\documentclass[aps,prl,twocolumn,superscriptaddress,floatfix,notitlepage]{revtex4-1}

\usepackage{amsthm}

\usepackage{graphicx,psfrag,color}
\usepackage{mathbbol,amsmath,amsfonts,amssymb,bm,dsfont,braket}
\usepackage{wasysym}
\usepackage[section]{placeins}
\usepackage{multirow}

\def\i{{\sf i}}
\newcommand{\ignore}[1]{ }

\usepackage{verbatim}

\usepackage{array}
\newcolumntype{x}[1]{>{\centering\hspace{0pt}}p{#1}}
\usepackage[normalem]{ulem}

\begin{document}

\title{Quantum Quenches of an SO(5) Pseudospin Reveal Higgs Bosons}

\begin{abstract}
Controlled dynamical probe measurement of complex order parameter fluctuations may reveal its massive collective excitations. Here, we design dynamical quench protocols to excite independently all ten midgap Higgs bosons in the isotropic Balian--Werthamer state of a spinfull $p$-wave superfluid or superconductor. The analysis is based on microscopic equations of motion of an SO(5) pseudospin, an extension of the usual Bloch equation to a five-dimensional space. Key to these protocols is the realization of quenches that break the rotational symmetry of the kinetic energy and exploit the irreducible representation of the angular momentum $J=2$. For perturbative quenches, we find ({\it non-decaying}) periodic oscillations in time of these Higgs modes. We present experimental protocols for superconductors (superfluids), with the intention of unveiling the nature of their order parameters.
\end{abstract}

\author{Qiao-Ru Xu}
\affiliation{\mbox{Department of Physics, Indiana University, Bloomington, Indiana 47405, USA}}
\author{Gerardo Ortiz}
\affiliation{\mbox{Department of Physics, Indiana University, Bloomington, Indiana 47405, USA}}
\affiliation{Indiana University Quantum Science and Engineering Center, Bloomington, Indiana 47408, USA}

\date{\today}
\maketitle

\begin{figure}[t]
\includegraphics[width=\columnwidth]{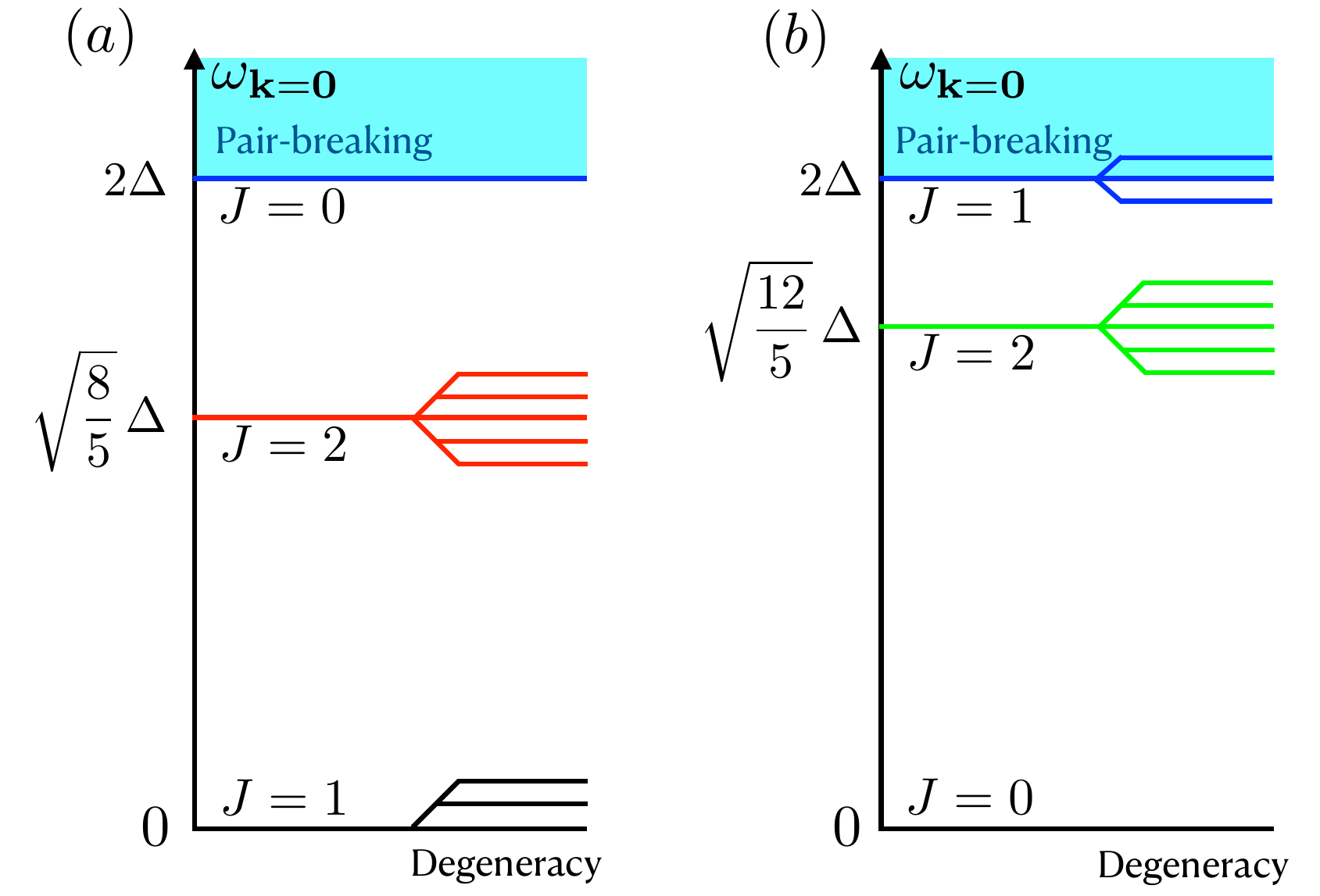}
\caption{Energy spectra of 18 collective excitations with zero momentum $\mathbf{k}$ in a neutral BW state (pair-breaking quasiparticle excitations appear above the gap $2 \Delta$). Panel $(a)$ is associated with the real part of the order parameter, with 3 degenerate Goldstone modes at zero energy (black lines), 5 degenerate Higgs modes at $\sqrt{8/5}\Delta$ (red lines) and 1 Higgs mode at $2\Delta$ (blue line). Panel $(b)$ is associated with the imaginary part of the order parameter, with 1 Goldstone mode at zero energy (black line), 5 degenerate Higgs modes at $\sqrt{12/5}\Delta$ (green lines) and 3 degenerate Higgs modes at $2\Delta$ (blue lines). Symmetry considerations support Nambu's partner relation in each angular momentum sector $J=0,1,2$: $[\omega_{(a)}^{(J)}]^2+[\omega_{(b)}^{(J)}]^2=4 \Delta^2$.}
\label{Fig_1}
\end{figure}

{\it Introduction.---}The discovery of a 125 GeV Higgs boson at LHC \cite{ATLAS,CMS} has generated renewed interest in studying Higgs physics in diverse physical systems \cite{Volovik2014,Varma2015}. In the condensed matter arena, in particular \cite{Varma2015}, Higgs bosons (or modes) are massive collective excitations associated with amplitude fluctuations of the order parameter associated to 
a particular broken symmetry phase of matter, such as neutral fermionic superfluids \cite{Wolfle1977,Volovik2016}, superconductors \cite{Varma2002,Shimano2020}, quantum magnets \cite{Ruegg2008, Jain2017,Hong2017}, or ultracold bosonic atoms in optical lattices \cite{Bloch2012,Pollet2012}. The simplest illustration is an $s$-wave superconductor with a U(1) order parameter, described by an Anderson SU(2) pseudospin \cite{Anderson1958}, whose transverse amplitude oscillation is related to only one Higgs mode at the continuous spectrum's minimum $2\Delta$. In contrast, a minimal model of an SO(5) pseudospin \cite{Hasegawa1979,Murakami1999} is needed to portray a spinful $p$-wave superfluid (either neutral or charged), whose order parameters belong to subgroups of the overall  SO$^{\,}_S$(3)$\times$SO$^{\,}_L$(3)$\times$U(1) symmetry \cite{Volovik2014}. When restricted to the isotropic Balian--Werthamer (BW) state \cite{BW,Sauls2022}, symmetry group classification leads to a total of 14 Higgs modes, 10 of them below $2\Delta$ (See Fig.\,\ref{Fig_1}).

To study dynamical properties of Higgs modes, the probing technique of dynamical quenches \cite{Mitra2018} has been widely exploited, especially in the case of $s$-wave \cite{Volkov1974,Levitov2004,Altshuler2006,Levitov2006,Yuzbashyan2006,Shimano2013,Shimano2014,Tsuji2015}, as well as spinless $p$-wave \cite{Foster2013} and $d$-wave \cite{Shimano2018,Schwarz2020} superconductors, mostly within the Anderson SU(2) pseudospin formalism. For example, it was found that for an $s$-wave superconductor, the only Higgs mode at $2\Delta$ displays a power law decay in time with oscillations after a perturbative quench of the interaction strength \cite{Volkov1974,Altshuler2006}, or the coupling to an external electromagnetic gauge field \cite{Tsuji2015}. Similar quench dynamics also applies to a spinless $p$-wave system \cite{Foster2013}. To the best of our knowledge, quench dynamics of Higgs modes in a spinful $p$-wave superfluid or superconductor has not been studied yet. In contrast to the 
$s$-wave superconductor, the rich structure of collective excitations and plethora of midgap Higgs modes below $2\Delta$ makes the case for quench dynamics in this system one of particular physical relevance. 
Natural questions include: What kind of quenches need to be designed to uncover those Higgs? What are physical manifestations of those bosons after quenches? And how can one realize those protocols experimentally? In this letter, we answer these questions within the SO(5) pseudospin formalism.

{\it Anisotropic Cooper Pairing.---}Consider the mean-field Hamiltonian of a spinful $p$-wave superfluid \cite{BW}
\begin{align}\label{H_MF}
\widehat{H}=
\sum_{\mathbf{k}s}\xi^{\,}_\mathbf{k}c^\dagger_{\mathbf{k}s} c^{\,}_{\mathbf{k}s} 
+\frac{1}{2}\sum_{\mathbf{k}ss'}(\Delta^{ss'}_\mathbf{k}c^\dagger_{\mathbf{k}s} c^\dagger_{-\mathbf{k}s'}+\text{H.c.}),
\end{align}
in three spatial dimensions, where $\xi^{\,}_\mathbf{k}=\frac{\hbar^2k^2}{2m}-\epsilon^{\,}_F$ is the kinetic energy of 
fermionic particles of momentum $\mathbf{k}$, mass $m$, and spin $s \in \{\uparrow,\downarrow\}$, measured from the Fermi energy $\epsilon^{\,}_F$, with creation (annihilation) 
operators $c^\dagger_{\mathbf{k}s}$ ($c^{\,}_{\mathbf{k}s}$). The  pairing $\Delta^{ss'}_\mathbf{k}=\sum_\mathbf{k'}
V^{\,}_{\mathbf{k}\mathbf{k'}}\braket{c^{\,}_{-\mathbf{k'}s'} c^{\,}_{\mathbf{k'}s}}=-\Delta^{s's}_\mathbf{-k}$ is an element of a $2\times2$ pairing matrix $\Delta^{\,}_\mathbf{k}$, where $V^{\,}_{\mathbf{k}\mathbf{k'}}\propto \mathbf{n}\cdot\mathbf{n'}$ describes two-particle interactions near the Fermi surface with $\mathbf{n}=\mathbf{k}/|\mathbf{k}|$. Note that $\Delta^{ss'}_\mathbf{k}$ will become time $t$ dependent and will be self-consistently determined later.

Equation\,(\ref{H_MF}) can be conveniently cast into the following form $\widehat{H}=\frac{1}{2}\sum_\mathbf{k}\hat{\Psi}^\dagger_{\mathbf{k}} H^{\,}_\text{BdG}
\hat{\Psi}^{\,}_{\mathbf{k}} + \sum_\mathbf{k}\xi^{\,}_\mathbf{k}$, in terms of the 4-component Nambu spinor $\hat{\Psi}^\dagger_{\mathbf{k}}=[c^\dagger_{\mathbf{k}\uparrow},c^\dagger_{\mathbf{k}\downarrow},c^{\,}_{\mathbf{-k}\downarrow},-c^{\,}_{\mathbf{-k}\uparrow}]$, where $H^{\,}_\text{BdG}=\xi^{\,}_\mathbf{k}\tau^3
+\mathbf{d}^{\,}_\mathbf{k}\cdot(\frac{1}{2}\tau^+\boldsymbol{\sigma})
+\mathbf{d}^*_\mathbf{k}\cdot(\frac{1}{2}\tau^-\boldsymbol{\sigma})$ is the Bogoliubov--de Gennes Hamiltonian, with Pauli matrices $\tau^j$ and $\sigma^j\,(j=1,2,3)$ acting on the Nambu and spin spaces, respectively, $\tau^\pm=\tau^1\pm \i\tau^2$, and the complex vector $\mathbf{d}^{\,}_\mathbf{k}$ is defined through the relation $-\i\Delta^{\,}_\mathbf{k}\sigma^2=\mathbf{d}^{\,}_\mathbf{k}\cdot\boldsymbol{\sigma}$ so that $\mathbf{d}^{\,}_\mathbf{k}=-\mathbf{d}^{\,}_\mathbf{-k}$ has odd parity for the spin-triplet pairing of spin-1/2 fermions. 

To avoid double counting in the Nambu basis, we define the domain ${\cal D}=\{\mathbf{k}|k^{\,}_1>0\}\cup\{\mathbf{k}|k^{\,}_2>0,k^{\,}_1=0\}\cup\{\mathbf{k}|k^{\,}_3>0,k^{\,}_2=0,k^{\,}_1=0\}$ hereinafter. Immediately, we see that the building blocks of $\widehat{H}$ are $U^j_\mathbf{k}=\frac{1}{2}\hat{\Psi}^\dagger_{\mathbf{k}}\tau^1\sigma^j\hat{\Psi}^{\,}_{\mathbf{k}}=-U^j_\mathbf{-k}$, $V^j_\mathbf{k}=\frac{1}{2}\hat{\Psi}^\dagger_{\mathbf{k}}\tau^2\sigma^j\hat{\Psi}^{\,}_{\mathbf{k}}=-V^j_\mathbf{-k}$ and $N^-_\mathbf{k}=\frac{1}{2}\hat{\Psi}^\dagger_{\mathbf{k}} \tau^3
\hat{\Psi}^{\,}_{\mathbf{k}}=N^-_\mathbf{-k}$, where $2(N^-_\mathbf{k}+1)=N^{\,}_\mathbf{k}$ is the number operator counting all spinful fermions with momenta $\pm\mathbf{k}$, and $\mp\frac{1}{2}(U^\pm_\mathbf{k} +\i V^\pm_\mathbf{k})=T^\dagger_{\pm 1\mathbf{k}}$ and $U^3_\mathbf{k} +\i V^3_\mathbf{k}=T^\dagger_{0\mathbf{k}}$ are the spin-triplet fermion pair ($\mathbf{k},-\mathbf{k}$) creation operators $T^\dagger_{m\mathbf{k}}$ with spin projection $m\in\{\pm 1$, 0\}, after introducing the notation $U^\pm_\mathbf{k}=U^1_\mathbf{k}\pm \i U^2_\mathbf{k}$ and $V^\pm_\mathbf{k}=V^1_\mathbf{k}\pm \i V^2_\mathbf{k}$.

{\it SO(5) Pseudospin.---}In order to establish the connection between building blocks 
\{$U^j_\mathbf{k},V^j_\mathbf{k},N^-_\mathbf{k}$\} of $\widehat{H}$ and an SO(5) pseudospin, we introduce the total spin operator at momenta $\pm\mathbf{k}$, $S^j_\mathbf{k}=\frac{1}{2}\hat{\Psi}^\dagger_{\mathbf{k}}\sigma^j\hat{\Psi}^{\,}_{\mathbf{k}}$, and define an antisymmetric rank-2 tensor $L^{\mu\nu}_\mathbf{k}=-L^{\nu\mu}_\mathbf{k}\,(\mu,\nu=1,2,3,4,5)$ through relations $L^{ij}_\mathbf{k}=\epsilon^{ijk}S^k_\mathbf{k}$, contracted with Levi-Civita tensor $\epsilon^{ijk}$, $L^{4j}_\mathbf{k}=U^j_\mathbf{k}$, $L^{5j}_\mathbf{k}=V^j_\mathbf{k}$ and $L^{45}_\mathbf{k}=N^-_\mathbf{k}$. One can verify that $L^{\mu\nu}_\mathbf{k}$ satisfy the SO(5) algebra
\begin{align}
\hspace*{-0.1cm}[L^{\alpha\beta}_\mathbf{k}, L^{\mu\nu}_\mathbf{k}]
\hspace*{-0.05cm}=\hspace*{-0.05cm}\i(\delta^{\alpha\mu}L^{\beta\nu}_\mathbf{k}\hspace*{-0.05cm}+\hspace*{-0.05cm}\delta^{\beta\nu}L^{\alpha\mu}_\mathbf{k}\hspace*{-0.05cm}-\hspace*{-0.05cm}\delta^{\alpha\nu}L^{\beta\mu}_\mathbf{k}\hspace*{-0.05cm}-\hspace*{-0.05cm}\delta^{\beta\mu}L^{\alpha\nu}_\mathbf{k}),\label{defso(5)}
\end{align}
while $[L^{\alpha\beta}_\mathbf{p}, L^{\mu\nu}_\mathbf{p'}]=0$ when $\mathbf{p}\neq\mathbf{p'}$ and $\mathbf{p},\,\mathbf{p'}\in {\cal D}\cup\{\mathbf{0}\}$. Explicitly, we have the SO(5) algebra of Eq.\,(\ref{defso(5)}) 
\begin{align}
&[S^i_\mathbf{k}, S^j_\mathbf{k}]=\i\epsilon^{ijk}S^k_\mathbf{k},\label{S}\\
[S^i_\mathbf{k}, U^j_\mathbf{k}]=&\i\epsilon^{ijk}U^k_\mathbf{k},\quad [U^i_\mathbf{k}, U^j_\mathbf{k}]=\i\epsilon^{ijk}S^k_\mathbf{k},\label{SU}\\
[S^i_\mathbf{k}, V^j_\mathbf{k}]=&\i\epsilon^{ijk}V^k_\mathbf{k},\quad [V^i_\mathbf{k}, V^j_\mathbf{k}]=\i\epsilon^{ijk}S^k_\mathbf{k},\label{SV}\\
\hspace*{-0.1cm}[U^j_\mathbf{k}, V^j_\mathbf{k}]=\i N^-_\mathbf{k},\quad &[V^j_\mathbf{k}, N^-_\mathbf{k}]=\i U^j_\mathbf{k},\quad[N^-_\mathbf{k}, U^j_\mathbf{k}]=\i V^j_\mathbf{k},\hspace*{-0.2cm}\label{UVN}
\end{align}
with either \{$\mathbf{S}^{\,}_\mathbf{k},\mathbf{U}^{\,}_\mathbf{k}$\} or \{$\mathbf{S}^{\,}_\mathbf{k},\mathbf{V}^{\,}_\mathbf{k}$\} forming an SO(4) subalgebra, \{$U^j_\mathbf{k},V^j_\mathbf{k},N^-_\mathbf{k}$\} an SO(3) subalgebra, and all other commutators vanishing. Note that when $\mathbf{p}=\mathbf{0}$ the 10 generators of SO(5) break down to the 4 generators \{$\frac{\mathbf{S}^{\,}_\mathbf{0}}{2},\frac{N^{\,}_\mathbf{0}}{2}$\} of U(2) while the other 6 vanish.

As an extension of the 3-dimensional Anderson SU(2) pseudospin \cite{Anderson1958}, we can arrange the above 10 generators of the SO(5) algebra as a 10-dimensional SO(5) pseudospin \cite{Hasegawa1979} $\mathbf{L}^{\,}_\mathbf{k}=[\mathbf{S}^{\,}_\mathbf{k}\, \mathbf{U}^{\,}_\mathbf{k}\, \mathbf{V}^{\,}_\mathbf{k}\, N^-_\mathbf{k}]^\text{T}$. It is necessary for a pseudospin to be
an Hermitian operator, and indeed one can check that we have $\mathbf{L}^{\,}_\mathbf{k}=\mathbf{L}^\dagger_\mathbf{k}$.

{\it Equations of Motion.---}To derive equations of motion for our system, one needs to first rewrite Hamiltonian $\widehat{H}$ in terms of the SO(5) pseudospin algebra
\begin{align}
\widehat{H}=\sideset{}{'}\sum_\mathbf{k}(\mathbf{H}^{\,}_\mathbf{k}\cdot\mathbf{L}^{\,}_\mathbf{k}+2\xi^{\,}_\mathbf{k})+\mathbf{H}^{\,}_\mathbf{0}\cdot\frac{\mathbf{L}^{\,}_\mathbf{0}}{2}+\xi^{\,}_\mathbf{0},\label{SO(5)MF-2}
\end{align}
where the primed summation is taken over the domain ${\cal D}$
and $\mathbf{H}^{\,}_\mathbf{k}=[\mathbf{0},\,2\text{Re\,}\mathbf{d}^*_\mathbf{k},\,2\text{Im\,}\mathbf{d}^*_\mathbf{k},\,2\xi^{\,}_\mathbf{k}]$ is a 10-dimensional pseudo-magnetic field, with its first 3 elements (i.e., the effective magnetic field acting on $\mathbf{S}^{\,}_\mathbf{k}$) vanishing. In general, if we define the 10 elements of the pseudo-magnetic field as $\mathbf{H}^{\,}_\mathbf{k}=[\mathbf{H}^S_\mathbf{k}\, \mathbf{H}^U_\mathbf{k}\, \mathbf{H}^V_\mathbf{k}\, H^N_\mathbf{k}]=\mathbf{H}^*_\mathbf{k}$, where
\begin{align}
&\mathbf{H}^U_\mathbf{k}=2\sideset{}{'}\sum_\mathbf{k'}
V^{\,}_{\mathbf{k}\mathbf{k'}}\braket{\mathbf{U}^{\,}_{\mathbf{k'}}}=2\text{Re\,}\mathbf{d}^*_\mathbf{k},\\
&\mathbf{H}^V_\mathbf{k}=2\sideset{}{'}\sum_\mathbf{k'}
V^{\,}_{\mathbf{k}\mathbf{k'}}\braket{\mathbf{V}^{\,}_{\mathbf{k'}}}=2\text{Im\,}\mathbf{d}^*_\mathbf{k},
\end{align}
then from Eqs.\,(\ref{S})-(\ref{SO(5)MF-2}) the equations of motion of the SO(5) pseudospin can be compactly written as follows
\begin{align}\label{e.o.m.}
\hspace*{-0.2cm}\frac{d}{dt}\hspace*{-0.05cm}\braket{\mathbf{L}^{\,}_\mathbf{k}}\hspace*{-0.05cm}=\hspace*{-0.05cm}\frac{1}{\hbar}\hspace*{-0.05cm}
\begin{bmatrix}
\mathbf{H}^S_\mathbf{k}\times &\hspace*{0.4cm}\mathbf{H}^U_\mathbf{k}\times &\hspace*{0.5cm}\mathbf{H}^V_\mathbf{k}\times &\mathbf{0}\\
\mathbf{H}^U_\mathbf{k}\times &\hspace*{0.4cm}\mathbf{H}^S_\mathbf{k}\times &-H^N_\mathbf{k} &\hspace*{0.3cm}\mathbf{H}^V_\mathbf{k}\\
\mathbf{H}^V_\mathbf{k}\times &\hspace*{0.2cm}H^N_\mathbf{k} &\hspace*{0.4cm}\mathbf{H}^S_\mathbf{k}\times &-\mathbf{H}^U_\mathbf{k}\\
\mathbf{0}\cdot &-\mathbf{H}^V_\mathbf{k}\cdot &\hspace*{0.3cm}\mathbf{H}^U_\mathbf{k}\cdot &0
\end{bmatrix}
\hspace*{-0.05cm}\braket{\mathbf{L}^{\,}_\mathbf{k}}\hspace*{-0.05cm},\hspace*{-0.2cm}
\end{align}
which is the analogue, and extension, of the Bloch equation $\frac{d}{dt}\braket{\frac{\mathbf{S}^{\,}_\mathbf{0}}{2}}=\frac{1}{\hbar}\mathbf{H}^S_\mathbf{0}\times\braket{\frac{\mathbf{S}^{\,}_\mathbf{0}}{2}}$ of the spin-1/2 operator $\frac{\mathbf{S}^{\,}_\mathbf{0}}{2}$. For example, the third and fourth lines of Eq.\,(\ref{e.o.m.}) read as  $\frac{d}{dt}\hspace*{-0.05cm}\braket{\mathbf{V}^{\,}_\mathbf{k}}\hspace*{-0.06cm}=\hspace*{-0.06cm}\frac{1}{\hbar}\hspace*{-0.05cm}\left[\mathbf{H}^V_\mathbf{k}\hspace*{-0.1cm}\times\hspace*{-0.1cm}\braket{\mathbf{S}^{\,}_\mathbf{k}}\hspace*{-0.05cm}+\hspace*{-0.05cm}H^N_\mathbf{k}\hspace*{-0.1cm}\braket{\mathbf{U}^{\,}_\mathbf{k}}\hspace*{-0.05cm}+\hspace*{-0.05cm}\mathbf{H}^S_\mathbf{k}\hspace*{-0.1cm}\times\hspace*{-0.1cm}\braket{\mathbf{V}^{\,}_\mathbf{k}}\hspace*{-0.05cm}-\hspace*{-0.05cm}\mathbf{H}^U_\mathbf{k}\hspace*{-0.1cm}\braket{N^-_\mathbf{k}}\right]$ and $\frac{d}{dt}\braket{N^-_\mathbf{k}}=\frac{1}{\hbar}\left[-\mathbf{H}^V_\mathbf{k}\cdot\braket{\mathbf{U}^{\,}_\mathbf{k}}+\mathbf{H}^U_\mathbf{k}\cdot\braket{\mathbf{V}^{\,}_\mathbf{k}}\right]$. Also, it is clear that $\frac{N^{\,}_\mathbf{0}}{2}$ is a constant of motion of the system.

{\it Quench Protocols.---}As mentioned before, there is a total of 14 Higgs modes associated to the BW state (see Fig.\,1). Among those, 10 midgap states of total angular momentum $J=2$ have energies $\sqrt{8/5}\Delta$ (5-fold) and $\sqrt{12/5}\Delta$ (5-fold) 
and are known as real and imaginary squashing modes, respectively. In practice, depending on initial conditions and quench protocol, real (imaginary) squashing modes are not necessarily associated with oscillations of the real (imaginary) part of the order parameter $\mathbf{H}^U_\mathbf{k}$ ($\mathbf{H}^V_\mathbf{k}$), but they can be associated with oscillations of the imaginary (real). For the sake of clarity, in what follows we set the initial condition of $\mathbf{H}^{\,}_\mathbf{k}$ as $\mathbf{H}^{\,}_{\mathbf{k},0}=[\mathbf{0},\, 2\Delta^{\,}_0\mathbf{n},\, \mathbf{0},\, 2\xi^{\,}_\mathbf{k}]$ so that, apart from 1 (3) Higgs mode(s) with $J=0$ ($J=1$) at $2\Delta$, 5 real (imaginary) squashing modes are associated with oscillations of $\mathbf{H}^U_\mathbf{k}$ ($\mathbf{H}^V_\mathbf{k}$) only if we choose a suitable quench protocol. Note that for the BW state, we are able to define $\mathbf{d}^{*}_\mathbf{k}=\Delta\mathbf{n}$, so that $\mathbf{H}^U_\mathbf{k}=2(\text{Re\,}\Delta)\mathbf{n}$ and $\mathbf{H}^V_\mathbf{k}=2(\text{Im\,}\Delta)\mathbf{n}$. The condition $\Delta \in \mathbb{R}$ at $t=0$ leads to the  $\mathbf{H}^{\,}_{\mathbf{k},0}$ defined above.

\begin{figure}[t]
	\includegraphics[width=0.46\textwidth]{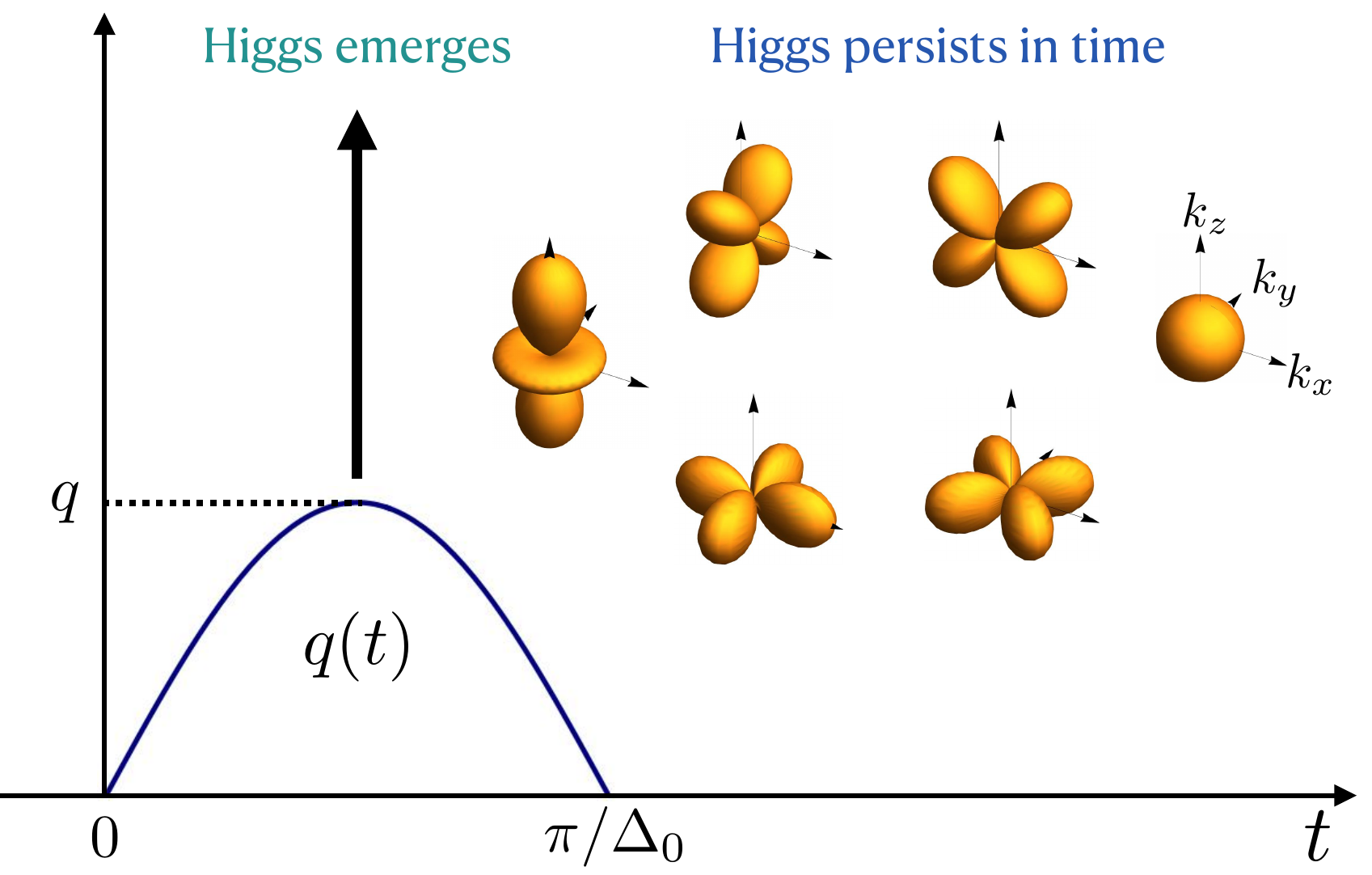}
	\caption{Quench protocols associated with $H^N_\mathbf{k}$. When $t\leq 0$ and $t>\pi/\Delta^{\,}_0$, $H^N_\mathbf{k}=2\xi^{\,}_\mathbf{k}$ is rotation symmetric in momentum space. When $0<t\leq \pi/\Delta^{\,}_0$, we set $H^N_\mathbf{k}=2\xi^{\,}_\mathbf{k}+q(t)\sum^{\,}_{ij}Q^{\,}_{ij}n^{\,}_in^{\,}_j\xi^{\,}_\mathbf{k}$ or $H^N_\mathbf{k}=2\xi^{\,}_\mathbf{k}+q(t)\sum^{\,}_{ij}Q^{\,}_{ij}n^{\,}_in^{\,}_j|\xi^{\,}_\mathbf{k}|$, where $q(t)$ and $\sum^{\,}_{ij}Q^{\,}_{ij}n^{\,}_in^{\,}_j$ can be chosen as $q \sin (\Delta^{\,}_0 \, t) \Theta(t)\Theta(\pi/\Delta^{\,}_0 -t)$ and $3d$ orbitals in momentum space, respectively, so as to generate the midgap Higgs modes. If $\sum^{\,}_{ij}Q^{\,}_{ij}n^{\,}_in^{\,}_j$ is chosen as the $1s$ orbital in momentum space, then only the $2\Delta^{\,}_0$ Higgs mode is excited.}
	\label{Fig2}
\end{figure}

Now we are ready to tackle the relevant quench protocols. Because Higgs modes with $J=2$ break the spherical symmetry of the order parameter associated to the BW state, quench protocols better break that symmetry as well. For example, quenching the interaction strength is the usual strategy to excite the Higgs mode in an SU(2) model. However, since this is an isotropic quench, only the Higgs mode with $J=0$ at $2\Delta$ can be excited in our SO(5) model. We checked that this is indeed the case. Another strategy is to couple the system to an external electromagnetic gauge field $\mathbf{A}(t)$. This strategy, which obviously is not designed for neutral superfluids, only induces a trivial phase factor of the $\mathbf{d}^{\,}_\mathbf{k}$ vector for charged ones, and therefore no Higgs mode is excited (see analytical solutions \cite{Suppl}). We will come back to this point later. Because there are spin components $\mathbf{S}^{\,}_\mathbf{k}$ of an SO(5) pseudospin $\mathbf{L}^{\,}_\mathbf{k}$, the third strategy is to couple the system to an external magnetic field $\mathbf{B}(t)$. Although a trivial coupling without magnetically polarizing the BW state cannot excite Higgs modes \cite{Suppl}, this strategy is still promising if we first magnetize the system a little bit and then couple it to $\mathbf{B}(t)$. While keeping an open mind to this possibility, we switch to a fourth strategy that is the focus of this paper by anisotropically quenching the kinetic energy, or equivalently the component $H^N_\mathbf{k}$ of the pseudo-magnetic field $\mathbf{H}^{\,}_\mathbf{k}$. As we will see, it provides a systematic way to excite independently those 10 massive Higgs modes.

To simplify the problem, the quench strength will be set such that, after the quench, we can still use the initial value of the gap $\Delta^{\,}_0$ rather than the value $\Delta^{\,}_{\infty}$ at $t \rightarrow \infty$ to characterize the Higgs modes (i.e., $|\Delta^{\,}_0-\Delta^{\,}_{\infty} | \ll \Delta_0$). Indeed, under a variety of circumstances, the system might not even reach a steady state as we will see below. Furthermore, if several Higgs modes get excited, then we will focus on dominant signals and neglect subdominant ones, several orders of magnitude smaller. To get to the point, as a quantum quench protocol, at $t=0^+$, we add a time-dependent perturbation $h^N_\mathbf{k}$ to $H^N_\mathbf{k}$ associated with the Hamiltonian $\widehat{H}$ of Eq.\,(\ref{SO(5)MF-2}), and denote
\begin{align}\label{DeviationFromEquil}
\hspace*{-0.2cm}\braket{\mathbf{L}^{\,}_\mathbf{k}}=\braket{\mathbf{L}^{\,}_\mathbf{k}}^{\,}_0+\mathbf{l}^{\,}_\mathbf{k}(t)\Theta(t),\,\,\,\,
\mathbf{H}^{\,}_\mathbf{k}=\mathbf{H}^{\,}_{\mathbf{k},0}+\mathbf{h}^{\,}_\mathbf{k}(t)\Theta(t),\hspace*{-0.2cm}
\end{align}
where we have $\braket{\mathbf{L}^{\,}_\mathbf{k}}^{\,}_0=[\braket{\mathbf{S}^{\,}_\mathbf{k}}^{\,}_0\, \braket{\mathbf{U}^{\,}_\mathbf{k}}^{\,}_0\, \braket{\mathbf{V}^{\,}_\mathbf{k}}^{\,}_0\, \braket{N^-_\mathbf{k}}^{\,}_0]^\text{T}$ the equilibrium value of $\braket{\mathbf{L}^{\,}_\mathbf{k}}$ at $t=0$,
$\mathbf{l}^{\,}_\mathbf{k}=[\mathbf{l}^S_\mathbf{k}\, \mathbf{l}^U_\mathbf{k}\, \mathbf{l}^V_\mathbf{k}\, l^N_\mathbf{k}]^\text{T}$ the deviation of $\braket{\mathbf{L}^{\,}_\mathbf{k}}$ from $\braket{\mathbf{L}^{\,}_\mathbf{k}}^{\,}_0$ when $t>0$, and $\Theta(t)$ the unit step function. Similar meanings are attributed to $\mathbf{H}^{\,}_{\mathbf{k},0}=[\mathbf{H}^S_{\mathbf{k},0}\, \mathbf{H}^U_{\mathbf{k},0}\, \mathbf{H}^V_{\mathbf{k},0}\, H^N_{\mathbf{k},0}]$ and
$\mathbf{h}^{\,}_\mathbf{k}=[\mathbf{h}^S_\mathbf{k}\, \mathbf{h}^U_\mathbf{k}\, \mathbf{h}^V_\mathbf{k}\, h^N_\mathbf{k}]$. As mentioned above, we set $\mathbf{H}^{\,}_{\mathbf{k},0}=[\mathbf{0},\, 2\Delta^{\,}_0\mathbf{n},\, \mathbf{0},\, 2\xi^{\,}_\mathbf{k}]$ as the initial condition and define $E^{\,}_{\mathbf{k},0}=\sqrt{\xi^2_\mathbf{k}+\Delta^2_0}$. Then we have $\braket{\mathbf{L}^{\,}_\mathbf{k}}^{\,}_0=[\mathbf{0},\, \frac{-\Delta^{\,}_0}{E^{\,}_{\mathbf{k},0}}\mathbf{n},\, \mathbf{0},\, \frac{-\xi^{\,}_\mathbf{k}}{E^{\,}_{\mathbf{k},0}}]^\text{T}$ antiparallel to $\mathbf{H}^{\,}_{\mathbf{k},0}$ in the ground state. If we define the fluctuations
\begin{align}
(\mathbf{h}^U_\mathbf{k})^{\,}_i=\sum^3_{j=1} h^U_{ij}n^{\,}_j,\quad (\mathbf{h}^V_\mathbf{k})^{\,}_i=\sum^3_{j=1} h^V_{ij}n^{\,}_j,
\end{align}
the 18 time-dependent fields $h^U_{ij}$ and $h^V_{ij}$ will be associated with the 18 collective modes of the order parameter, i.e., 14 Higgs modes and 4 Goldstone modes (see Fig.\,1). 

We next introduce two kinds of quench protocols 
\begin{align}
&h^N_\mathbf{k}=q(t)\sum^{\,}_{ij}Q^{\,}_{ij}n^{\,}_in^{\,}_j\xi^{\,}_\mathbf{k},\label{Quench1}\\
&h^N_\mathbf{k}=q(t)\sum^{\,}_{ij}Q^{\,}_{ij}n^{\,}_in^{\,}_j|\xi^{\,}_\mathbf{k}|,\label{Quench2}
\end{align}
with strength $q(t)$ [in the following we will use $q(t)=q \sin (\Delta^{\,}_0\, t) \Theta(t)\Theta(\pi/\Delta^{\,}_0 -t)$] and $Q$ a second-rank time-independent quench tensor. We claim that all 5 real (imaginary) squashing modes associated with $h^U_{ij}$ ($h^V_{ij}$) can be excited by the quench protocol of Eq.\,(\ref{Quench1}) (Eq.\,(\ref{Quench2})), and if only one of them is excited, then there is a one-to-one correspondence between the tensor $Q$ and the Higgs mode $\sum^{\,}_{ij}Q^{\,}_{ij}h^U_{ij}$ ($\sum^{\,}_{ij}Q^{\,}_{ij}h^V_{ij}$). More specifically, each squashing mode can be excited independently if $Q$ is chosen to be one of the following tensors
\begin{align}
\label{Q}
& \hspace*{-1.19cm}Q^{\,}_{\hat{x}^2+\hat{y}^2-2\hat{z}^2}=\begin{bmatrix}
1 &0 &0\\
0 &1 &0\\
0 &0 &-2
\end{bmatrix}, \ \  
Q^{\,}_{\hat{x}^2-\hat{y}^2}=\begin{bmatrix}
1 &0 &0\\
0 &-1 &0\\
0 &0 &0
\end{bmatrix}, \\
Q^{\,}_{\hat{x}\hat{y}}=&\begin{bmatrix}
0 &1 &0\\
1 &0 &0\\
0 &0 &0
\end{bmatrix}, \ 
Q^{\,}_{\hat{x}\hat{z}}=\begin{bmatrix}
0 &0 &1\\
0 &0 &0\\
1 &0 &0
\end{bmatrix}, \
Q^{\,}_{\hat{y}\hat{z}}=\begin{bmatrix}
0 &0 &0\\
0 &0 &1\\
0 &1 &0
\end{bmatrix}, \nonumber
\end{align}
which are in correspondence with the $3d$ orbitals of atomic physics $3d^{\,}_{z^2}$, $3d^{\,}_{x^2-y^2}$, $3d^{\,}_{xy}$, $3d^{\,}_{xz}$, and $3d^{\,}_{yz}$, i.e., the irreducible representations of 
$J=2$ (see Fig. \ref{Fig2}).

\begin{figure}[t]
	\includegraphics[width=\columnwidth]{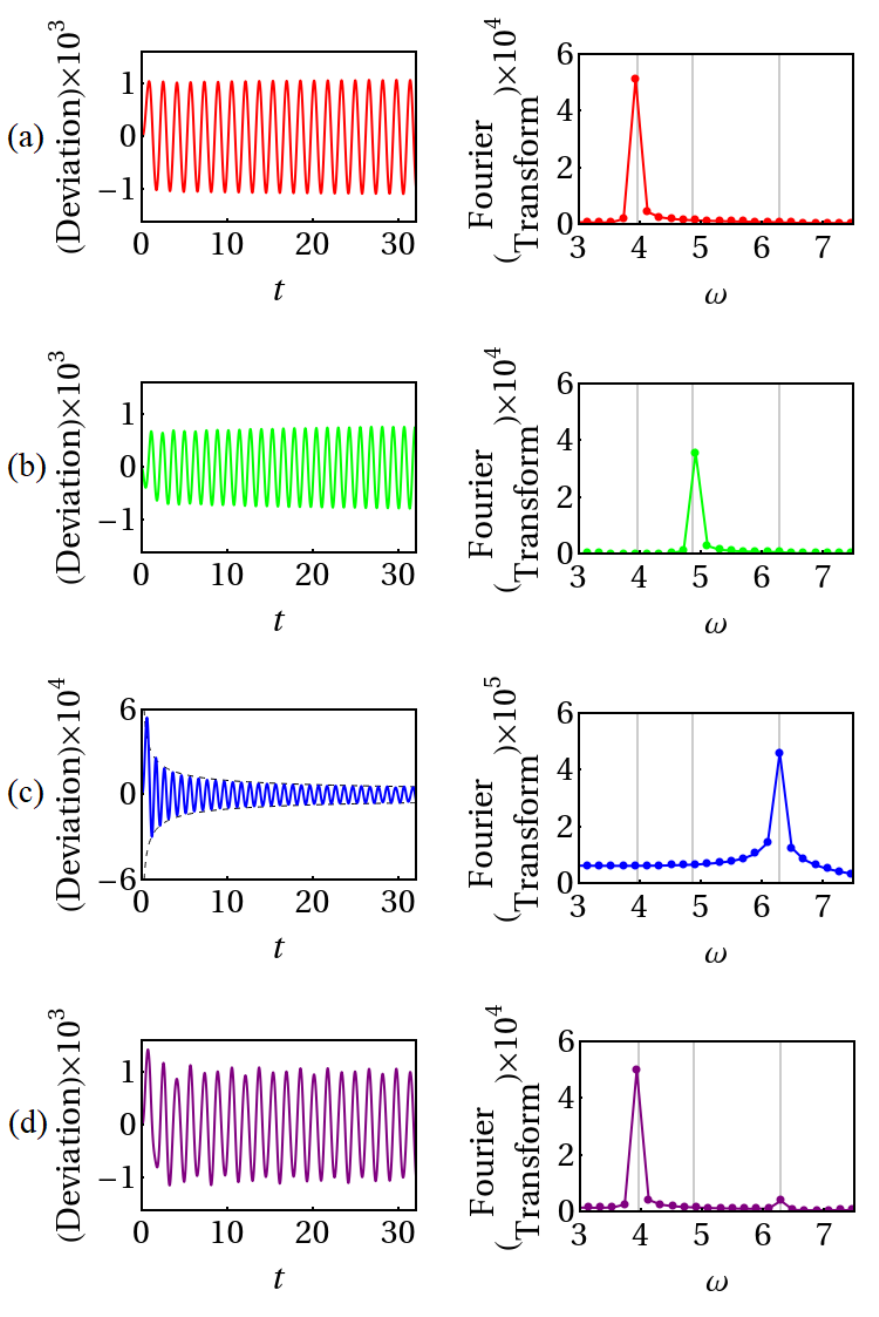}
	\caption{Quench dynamics of Higgs bosons, with parameters $\Delta^{\,}_0=\pi$ and $q(t)=10^{-4}\sin(\pi t) \Theta(t)\Theta(1-t)$. (a) Left: The real squashing mode $\sum^{\,}_{ij}(Q^{\,}_{\hat{x}^2+\hat{y}^2-2\hat{z}^2})_{ij}h^U_{ij}$ is plotted as a function of the time $t$ after a perturbative quench. Right: The discrete Fourier transform of the left is plotted as a function of the discrete frequency $\omega=2\pi(s-1)/32$ ($s\in\mathds{Z}$), with three vertical grid lines indicating Higgs modes at $\sqrt{8/5}\Delta^{\,}_0$, $\sqrt{12/5}\Delta^{\,}_0$ and $2\Delta^{\,}_0$, respectively. (b) and (c) are similar plots for the imaginary squashing mode $\sum^{\,}_{ij}(Q^{\,}_{\hat{x}^2+\hat{y}^2-2\hat{z}^2})_{ij}h^V_{ij}$ and the Higgs mode $\sum^{\,}_{ij}\delta^{\,}_{ij}h^U_{ij}$ with $J=0$ at $2\Delta$, respectively. Note that the two dashed curves $\propto \pm1/\sqrt{t}$ in the left figure of (c) mark the power-law decay of the Higgs. (d) is the plot of $\sum^{\,}_{ij}3\delta^{\,}_{i3}\delta^{\,}_{3j}h^U_{ij}$, indicating the simultaneous excitation of Higgs modes at both $\sqrt{8/5}\Delta^{\,}_0$ and $2\Delta^{\,}_0$.}
	\label{SO5Dynamics}
\end{figure}

{\it Higgs Bosons Dynamics.---}As mentioned, after a perturbative quench, the Higgs mode with $J=0$ at $2\Delta$ displays oscillations with a power law decay in an SU(2) model. This is also true for our SO(5) model. However, for the 10 midgap squashing modes ($J=2$), their dynamical behavior differs substantially. Since each of the 5 real (imaginary) squashing modes has qualitatively the same behavior, we next investigate numerically a representative in full detail. We set $\xi^{\,}_\mathbf{k}=\xi/8$ with $\xi\in[-36,36]$ an integer. For each $\xi$, there are 36 chosen $\mathbf{k}$-points and, therefore, there is a total of $73\times36=2628$ points sampled in  momentum space. Since there are 10 coupled equations for a fixed $\mathbf{k}$ in Eq.\,(\ref{e.o.m.}), we finally need to solve 26280 coupled equations of motion numerically. As to the other parameters, we choose $\Delta^{\,}_0=\pi$ and $q(t)=10^{-4}\sin(\pi t) \Theta(t)\Theta(1-t)$. As a result, we plot the quench dynamics of the real squashing mode $\sum^{\,}_{ij}(Q^{\,}_{\hat{x}^2+\hat{y}^2-2\hat{z}^2})_{ij}h^U_{ij}\equiv D(t)$ in Fig.\,\ref{SO5Dynamics} (a). As one can see, it oscillates periodically with no decay. In order to identify the frequency, we perform a discrete Fourier transform $\frac{1}{320}\sum_{\ell=1}^{320}D(t^{\,}_{\ell-1})e^{\i\omega t^{\,}_{\ell-1}}$ of $D(t)$, where $t^{\,}_{\ell-1}=0.1(\ell-1)$ and $\omega=2\pi(s-1)/32$ with $s$ integers. It is evident that the frequency peaks around the value of $\sqrt{8/5}\Delta^{\,}_0$. A similar oscillation behavior is also obtained for the imaginary squashing mode $\sum^{\,}_{ij}(Q^{\,}_{\hat{x}^2+\hat{y}^2-2\hat{z}^2})_{ij}h^V_{ij}$ in Fig.\,\ref{SO5Dynamics} (b), but with a larger frequency peaked around $\sqrt{12/5}\Delta^{\,}_0$. While damping of the  $2\Delta$ Higgs modes results from collisionless dephasing \cite{Volkov1974}, the lack of coupling between midgap Higgs modes and the continuum of quasiparticle excitations is responsible for the undamped oscillations.

For comparison, we plot the quench dynamics of the Higgs mode $\sum^{\,}_{ij}\delta^{\,}_{ij}h^U_{ij}$ with $J=0$ at $2\Delta$ in Fig.\,\ref{SO5Dynamics} (c), where the two dashed envelop curves $\pm3.2/\sqrt{t}$ in the left panel indicate the power-law decay as we already know in the usual $s$-wave superconductors. It is worth to mention that, since the identity $\mathds{1}^{\,}_3$ and the 5 symmetric quench tensors of Eq.\,(\ref{Q}) are linearly independent, any linear combination of the identity and other symmetric tensors as input for the quench tensor $Q$ of Eq.\,(\ref{Quench1}) will lead to two frequency peaks simultaneously. For example, if $h^N_\mathbf{k}=3q(t)n^2_3\xi^{\,}_\mathbf{k}$, i.e., $Q^{\,}_{ij}=3\delta^{\,}_{i3}\delta^{\,}_{3j}$ and $Q=\mathds{1}^{\,}_3-Q^{\,}_{\hat{x}^2+\hat{y}^2-2\hat{z}^2}$, we have the quench dynamics of $\sum^{\,}_{ij}3\delta^{\,}_{i3}\delta^{\,}_{3j}h^U_{ij}$ plotted in Fig.\,\ref{SO5Dynamics} (d), where two types of Higgs modes are excited simultaneously as expected.

{\it Experimental Realizations.---}We next propose experimental quench protocols for either a $p$-wave superfluid or a $p$-wave superconductor in the BW state. For a superfluid, the quench protocol of Eq.\,(\ref{Quench1}) can be achieved by quenching the mass tensor $M$ of the system, i.e., we set $h^N_\mathbf{k}=\hbar^2(k^2-k^2_F)\sum^{\,}_{ij}\left[\delta( M^{-1})^{\,}_{ij}\right]n^{\,}_in^{\,}_j$, which is equivalent to $2m\sum^{\,}_{ij}\left[\delta( M^{-1})^{\,}_{ij}\right]n^{\,}_in^{\,}_j\xi^{\,}_\mathbf{k}$ considering the fact that $2\xi^{\,}_\mathbf{k}=\hbar^2(k^2-k^2_F)/m$, with $k^{\,}_F$ the Fermi momentum. By comparison to  Eq.\,(\ref{Quench1}), we immediately see that $q(t)Q^{\,}_{ij}=2m\left[\delta( M^{-1})^{\,}_{ij}\right]$. It is worth mentioning that in ultrasound attenuation experiments  \cite{Lee1980,Halperin1980}, the coupling of the system to (polarized) phonons can be probably thought of as effectively quenching the mass tensor of our SO(5) model. For a superconductor, if we consider anisotropic band structures rather than the quadratic in momentum considered above, a quench protocol similar to that of Eq.\,(\ref{Quench1}) can be achieved by coupling the system to an electromagnetic gauge field $\mathbf{A}(t)$. For example, if $\xi^{\,}_\mathbf{k}=\xi^{\,}_\mathbf{-k}$, one can expand $H^N_\mathbf{k}=\xi^{\,}_{\mathbf{k}-{\sf q}\mathbf{A}/\hbar}+\xi^{\,}_{-\mathbf{k}-{\sf q}\mathbf{A}/\hbar}$ up to the second order in the gauge field as follows $H^N_\mathbf{k}=2\xi^{\,}_\mathbf{k}+\frac{{\sf q}^2}{\hbar^2}\sum^{\,}_{ij}A^{\,}_iA^{\,}_j\partial^{\,}_i\partial^{\,}_j\xi^{\,}_\mathbf{k}$, where the second term plays a role similar to Eq.\,(\ref{Quench1}), and by tuning the polarization of the gauge field, one can achieve the goal of quenching
the kinetic energy anisotropically or, in other words, quenching the effective mass tensor (by changing the hopping constants in different directions). Terahertz pump-probe spectroscopy \cite{Shimano2013,Shimano2014,Shimano2018} is extremely suitable for such experiments. For example, considering the band structure $\xi^{\,}_\mathbf{k}=-2\epsilon\sum^{3}_{j=1}\cos(ak^{\,}_j)-\epsilon^{\,}_F$, with $a$ the lattice constant and $\epsilon$ the hopping constant, one can check that $\frac{{\sf q}^2}{\hbar^2}\sum^{\,}_{ij}A^{\,}_iA^{\,}_j\partial^{\,}_i\partial^{\,}_j\xi^{\,}_\mathbf{k}=-2\epsilon\sum^3_{j=1}(-\frac{a^2{\sf q}^2}{\hbar^2}A^{2}_j)\cos(ak^{\,}_j)$, which is a special case of Eq.\,(\ref{Quench1}) with $q(t)Q^{\,}_{ij}=-\frac{a^2{\sf q}^2}{\hbar^2}A^{\,}_iA^{\,}_j\delta^{\,}_{ij}$ and $q(t)Q=-\frac{a^2{\sf q}^2|\mathbf{A}|^2}{3\hbar^2}\left[\mathds{1}^{\,}_3+\frac{A^2_1+A^2_2-2A^2_3}{2|\mathbf{A}|^2}Q^{\,}_{\hat{x}^2+\hat{y}^2-2\hat{z}^2}+\frac{3(A^2_1-A^2_2)}{2|\mathbf{A}|^2}Q^{\,}_{\hat{x}^2-\hat{y}^2}\right]$. If the gauge field is polarized as $\mathbf{A}=(0,0,A)$, then we have $q(t)Q=-\frac{a^2{\sf q}^2A^2}{3\hbar^2}\left(\mathds{1}^{\,}_3-Q^{\,}_{\hat{x}^2+\hat{y}^2-2\hat{z}^2}\right)$, which is the case we analyzed in Fig.\,\ref{SO5Dynamics} (d). We argue that the heavy-fermion superconductor UBe$^{\,}_{13}$ constitutes an excellent platform to conduct such experiments given the apparent controversy over the nature of its 
order parameter \cite{Ott1984,Shimizu2015,Shimizu2019}. 

{\it Concluding Remarks.---}
Collective excitations of the order parameter such as Higgs bosons encode essential information about the associated broken symmetry phase. In this letter we investigated the dynamical signature of midgap Higgs bosons in a spinful $p$-wave superfluid or superconductor after perturbative quenches of the kinetic energy, with a focus on quenching the (effective) mass tensor. Unlike oscillations with power-law decay of the Higgs boson in $s$-wave superconductors \cite{Shimano2013}, we found that midgap Higgs bosons display periodic oscillations with no decay, a signature that might be experimentally observable in the superfluid $^3$He-B or some candidate BW superconductors like UBe$^{\,}_{13}$. More generally, our experimental quench protocols provide smoking gun tests for unconventional superconducting/superfluid order parameters by revealing the dynamic behavior of the underlying Higgs excitations. It is known that the BW state, as a topological phase, also hosts topological quasiparticle excitations such as Majorana fermions \cite{JPCM,JPSJ}. It would be interesting to study quench dynamics of Majorana fermions in the BW state with a physical boundary. Finally it is also possible to extend the current SO(5) pseudospin formalism to study quench dynamics of Higgs bosons and Majorana fermions in the Anderson--Brinkman--Morel state \cite{AM,AB,Will} and in spin-3/2 cold atom systems \cite{Wu2003,Wu2006}.

{\it Acknowledgements.---}
The authors would like to thank G. E. Volovik for insightful discussions on the Higgs dynamics of an SU(2) pseudospin and its potential extensions. G.O. acknowledges support from the US Department of Energy grant DE-SC0020343.

\end{document}


\title{Supplemental Material for \\``Quantum Quenches of an SO(5) Pseudospin Reveal Higgs Bosons"}

\author{Qiao-Ru Xu}
\affiliation{\mbox{Department of Physics, Indiana University, Bloomington, Indiana 47405, USA}}
\author{Gerardo Ortiz}
\affiliation{\mbox{Department of Physics, Indiana University, Bloomington, Indiana 47405, USA}}
\affiliation{Indiana University Quantum Science and Engineering Center, Bloomington, Indiana 47408, USA}


\date{\today}
\maketitle

\section{Quench by coupling to an electromagnetic gauge field $\mathbf{A}(t)$}\label{A(t)}

As in the main text, we set $\mathbf{H}^{\,}_{\mathbf{k},0}=[\mathbf{0},\, 2\Delta^{\,}_0\mathbf{n},\, \mathbf{0},\, 2\xi^{\,}_\mathbf{k}]$ and define $E^{\,}_{\mathbf{k},0}=\sqrt{\xi^2_\mathbf{k}+\Delta^2_0}$, then we have $\braket{\mathbf{L}^{\,}_\mathbf{k}}^{\,}_0=[\mathbf{0},\, \frac{-\Delta^{\,}_0}{E^{\,}_{\mathbf{k},0}}\mathbf{n},\, \mathbf{0},\, \frac{-\xi^{\,}_\mathbf{k}}{E^{\,}_{\mathbf{k},0}}]^\text{T}$. After coupling to a gauge field $\mathbf{A}(t)$, the equation of motion of $\braket{\mathbf{L}^{\,}_\mathbf{k}}$ becomes
\begin{align}\label{e.o.m.A(t)}
\hspace*{-0.2cm}\frac{d}{dt}\hspace*{-0.05cm}\braket{\mathbf{L}^{\,}_\mathbf{k}}\hspace*{-0.05cm}=\hspace*{-0.05cm}
\begin{bmatrix}
0 &\hspace*{0.4cm}\mathbf{H}^U_\mathbf{k}\times &\hspace*{0.5cm}\mathbf{H}^V_\mathbf{k}\times &\mathbf{0}\\
\mathbf{H}^U_\mathbf{k}\times &\hspace*{0.4cm}0 &-H^N_\mathbf{k} &\hspace*{0.3cm}\mathbf{H}^V_\mathbf{k}\\
\mathbf{H}^V_\mathbf{k}\times &\hspace*{0.2cm}H^N_\mathbf{k} &\hspace*{0.4cm}0 &-\mathbf{H}^U_\mathbf{k}\\
\mathbf{0}\cdot &-\mathbf{H}^V_\mathbf{k}\cdot &\hspace*{0.3cm}\mathbf{H}^U_\mathbf{k}\cdot &0
\end{bmatrix}
\hspace*{-0.05cm}\braket{\mathbf{L}^{\,}_\mathbf{k}}\hspace*{-0.05cm},\hspace*{-0.2cm}
\end{align}
where we have set $\hbar=1$ and $H^N_\mathbf{k}=\xi^{\,}_{\mathbf{k}-{\sf q}\mathbf{A}}+\xi^{\,}_{-\mathbf{k}-{\sf q}\mathbf{A}}$. Using $\xi^{\,}_\mathbf{k}=\frac{k^2}{2m}-\epsilon^{\,}_F$, we have $H^N_\mathbf{k}=2\xi^{\,}_\mathbf{k}+\frac{{\sf q}^2}{m}\mathbf{A}^2$. The exact analytic solution is given by
\begin{align}
&\braket{\mathbf{S}^{\,}_\mathbf{k}}=\mathbf{0},\quad\braket{N^{-}_\mathbf{k}}=\frac{-\xi^{\,}_\mathbf{k}}{E^{\,}_{\mathbf{k},0}},\\
&\braket{\mathbf{U}^{\,}_\mathbf{k}}=\frac{-\Delta^{\,}_0}{E^{\,}_{\mathbf{k},0}}\cos\left(\frac{{\sf q}^2}{m}\int^t_0 dt'\mathbf{A}^2(t')\right)\mathbf{n},\\
&\braket{\mathbf{V}^{\,}_\mathbf{k}}=\frac{-\Delta^{\,}_0}{E^{\,}_{\mathbf{k},0}}\sin\left(\frac{{\sf q}^2}{m}\int^t_0 dt'\mathbf{A}^2(t')\right)\mathbf{n},\\
&\mathbf{H}^{U}_\mathbf{k}=2\Delta^{\,}_0\cos\left(\frac{{\sf q}^2}{m}\int^t_0 dt'\mathbf{A}^2(t')\right)\mathbf{n},\\
&\mathbf{H}^{V}_\mathbf{k}=2\Delta^{\,}_0\sin\left(\frac{{\sf q}^2}{m}\int^t_0 dt'\mathbf{A}^2(t')\right)\mathbf{n},\\
&\mathbf{d}^{\,}_\mathbf{k}=\left(\Delta^{\,}_0e^{-\i\frac{{\sf q}^2}{m}\int^t_0 dt'\mathbf{A}^2(t')}\right)\mathbf{n}.\label{SolnforA}
\end{align}
From Eq.\,(\ref{SolnforA}), we see that after the quench only a trivial phase factor $e^{-\i\frac{{\sf q}^2}{m}\int^t_0 dt'\mathbf{A}^2(t')}$ is induced for the $\mathbf{d}^{\,}_\mathbf{k}$ vector and, therefore, no Higgs modes can be excited. However, by coupling to an electromagnetic gauge field, we see in the main text that Higgs modes indeed can be excited in $p$-wave superconductors that have anisotropic band structures rather than the quadratic one considered here.

\section{Quench by coupling to an external magnetic field $\mathbf{B}(t)$}\label{B(t)}

As in the main text, we set $\mathbf{H}^{\,}_{\mathbf{k},0}=[\mathbf{0},\, 2\Delta^{\,}_0\mathbf{n},\, \mathbf{0},\, 2\xi^{\,}_\mathbf{k}]$ and define $E^{\,}_{\mathbf{k},0}=\sqrt{\xi^2_\mathbf{k}+\Delta^2_0}$, then we have $\braket{\mathbf{L}^{\,}_\mathbf{k}}^{\,}_0=[\mathbf{0},\, \frac{-\Delta^{\,}_0}{E^{\,}_{\mathbf{k},0}}\mathbf{n},\, \mathbf{0},\, \frac{-\xi^{\,}_\mathbf{k}}{E^{\,}_{\mathbf{k},0}}]^\text{T}$. After coupling to a magnetic filed $\mathbf{B}(t)$, the equation of motion of $\braket{\mathbf{L}^{\,}_\mathbf{k}}$ is
\begin{align}
\hspace*{-0.2cm}\frac{d}{dt}\hspace*{-0.05cm}\braket{\mathbf{L}^{\,}_\mathbf{k}}\hspace*{-0.05cm}=\hspace*{-0.05cm}
\begin{bmatrix}
\mathbf{B}\times &\hspace*{0.4cm}\mathbf{H}^U_\mathbf{k}\times &\hspace*{0.5cm}\mathbf{H}^V_\mathbf{k}\times &\mathbf{0}\\
\mathbf{H}^U_\mathbf{k}\times &\hspace*{0.4cm}\mathbf{B}\times &-H^N_\mathbf{k} &\hspace*{0.3cm}\mathbf{H}^V_\mathbf{k}\\
\mathbf{H}^V_\mathbf{k}\times &\hspace*{0.2cm}H^N_\mathbf{k} &\hspace*{0.4cm}\mathbf{B}\times &-\mathbf{H}^U_\mathbf{k}\\
\mathbf{0}\cdot &-\mathbf{H}^V_\mathbf{k}\cdot &\hspace*{0.3cm}\mathbf{H}^U_\mathbf{k}\cdot &0
\end{bmatrix}
\hspace*{-0.05cm}\braket{\mathbf{L}^{\,}_\mathbf{k}}\hspace*{-0.05cm},\hspace*{-0.2cm}
\end{align}
where we have set $\hbar=1$. The exact solution to the above equation is as follows
\begin{align}
\braket{\mathbf{S}^{\,}_\mathbf{k}}=\braket{\mathbf{V}^{\,}_\mathbf{k}}=\mathbf{H}^{V}_\mathbf{k}&=\mathbf{0},\quad
\braket{N^{-}_\mathbf{k}}=\frac{-\xi^{\,}_\mathbf{k}}{E^{\,}_{\mathbf{k},0}},\\
\braket{\mathbf{U}^{\,}_\mathbf{k}}\hspace*{-0.05cm}=\hspace*{-0.05cm}R\braket{\mathbf{U}^{\,}_\mathbf{k}}^{\,}_0,\quad
\mathbf{H}^{U}_\mathbf{k}&=2\Delta^{\,}_0R\mathbf{n},\quad
\mathbf{d}^{\,}_\mathbf{k}=\Delta^{\,}_0R\mathbf{n},\label{SolnforB}
\end{align}  
where $R$ is a time-dependent rotation matrix such that $\braket{\mathbf{U}^{\,}_\mathbf{k}}\hspace*{-0.05cm}=\hspace*{-0.05cm}R\braket{\mathbf{U}^{\,}_\mathbf{k}}^{\,}_0$ is a solution to the following equation
\begin{align}\label{BlochEq}
\frac{d}{dt}\braket{\mathbf{U}^{\,}_\mathbf{k}}=\mathbf{B}\times\braket{\mathbf{U}^{\,}_\mathbf{k}}.
\end{align}
From Eq.\,(\ref{SolnforB}), we see that the $\mathbf{d}^{\,}_\mathbf{k}$ vector belongs to the rotationally degenerate BW ground states and, therefore, no Higgs modes can be excited. Nonetheless, it is worth mentioning that when we choose the magnetic field as  $\mathbf{B}=\omega^{\,}_0\hat{z}+\omega^{\,}_1\big[\cos(\omega t)\hat{x}+\sin(\omega t)\hat{y}\big]$, Eq.\,(\ref{BlochEq}) has a compact solution with $R=R^{\,}_3R^{\,}_2R^{\,}_1R^{-1}_2$, where
\begin{align}
&R^{\,}_1=\begin{bmatrix}
1 						 		&0 					&0\\
0 						 		&\cos(\Omega t) 	&\sin(\Omega t)\\
0 						 		&-\sin(\Omega t) 	&\cos(\Omega t)
\end{bmatrix},\,
R^{\,}_2=\begin{bmatrix}
-\frac{\omega_1}{\Omega} 		&0 					&-\frac{\omega-\omega_0}{\Omega}\\
0 				 		 		&1 					&0\\
\frac{\omega-\omega_0}{\Omega} 	&0 					&-\frac{\omega_1}{\Omega}
\end{bmatrix},\nonumber\\
&R^{\,}_3=\begin{bmatrix}
\cos(\omega t) 		 			&-\sin(\omega t) 	&0\\
\sin(\omega t) 		 			&\cos(\omega t) 	&0\\
0 					 			&0 					&1
\end{bmatrix},\,\,\,\,
\Omega=\sqrt{(\omega-\omega_0)^2+\omega^2_1}.\nonumber
\end{align}
The above analytic solution is a generalization of the problem of Rabi oscillations of a spin-1/2 in a magnetic field $\mathbf{B}$ to a three-component vector.